\documentclass[aps,pra,twocolumn,superscriptaddress,floatfix,
nofootinbib,showpacs,longbibliography]{revtex4-2}

\usepackage[normalem]{ulem}
\usepackage{braket}
\usepackage{xcolor}
\usepackage[utf8]{inputenc}
\usepackage[T1]{fontenc} 
\usepackage[british]{babel} 
\usepackage[sc,osf]{mathpazo}
\usepackage[scaled=0.86]{berasans} 
\usepackage[colorlinks=true, citecolor=blue, urlcolor=blue]{hyperref} 
\usepackage{physics}
\usepackage{graphicx} 
\usepackage[babel]{microtype} 
\usepackage{amsmath,amssymb,amsthm,bm,amsfonts,mathrsfs,bbm} 

\usepackage{xspace} 
\definecolor{wrwrwr}{rgb}{0.3803921568627451,0.3803921568627451,0.3803921568627451}
\definecolor{rvwvcq}{rgb}{0.08235294117647059,0.396078431372549,0.7529411764705882}
\usepackage{pgf,tikz,pgfplots}
\usetikzlibrary{calc}
\pgfplotsset{compat=1.15}
\usepackage{mathrsfs}
\usetikzlibrary{arrows,snakes}
\usetikzlibrary{positioning}
\usepackage{xcolor}
\usepackage{appendix}
\usepackage{multirow}
\usepackage{array}
\usepackage{bigstrut}
\usepackage{braket}
\usepackage{color}
\usepackage{natbib}
\usepackage{multirow}
\usepackage{float}
\usepackage[caption = false]{subfig}
\usepackage{xcolor,colortbl}
\usepackage{color}

\newtheorem{theorem}{Theorem}

\newcommand{\proj}[1]{\ket{#1}\bra{#1}}

\begin{document}

	\title{Network-assist free self-testing of genuine multipartite entangled states}
	
	\author{Ranendu Adhikary}
	\affiliation{Electronics and Communication Sciences Unit, Indian Statistical Institute, 203 B.T. Road, Kolkata 700108, India.}
	\email{ronjumath@gmail.com}
	
	\author{Abhishek Mishra}
	\affiliation{Laboratoire d'Information Quantique, Université libre de Bruxelles, Belgium.}
	\email{abhishek.mishra@ulb.be}
	
	\author{Ramij Rahaman}
	\affiliation{Physics and Applied Mathematics Unit, Indian Statistical Institute, 203 B.T. Road, Kolkata 700108, India.}
	\email{ramijrahaman@isical.ac.in}
	
\begin{abstract}
{Self-testing is a method to certify quantum states and measurements in a device-independent way. The device-independent certification of quantum properties is purely based on input-output measurement statistics of the involved devices with minimal knowledge about their internal workings. Bipartite pure entangled states can be self-tested, but, in the case of multipartite pure entangled states, the answer is not so straightforward. Nevertheless, \v{S}upi\'{c} et al. \cite{vsupic2023quantum} recently introduced a novel self-testing method for any pure entangled quantum state, which leverages network assistance and relies on bipartite entangled measurements. Hence, their scheme loses the true device-independent flavor of self-testing. In this regard, we provide a self-testing scheme for genuine multipartite pure entangle states in the true sense by employing a generalized Hardy-type non-local argument. 
	Our scheme involves only local operations and classical communications and does not depend on bipartite entangled measurements and is free from any network assistance. In addition, we provide the device-independent bound of the maximum probability of success for generalized Hardy-type nonlocality argument.}

\end{abstract}
\maketitle

\section{Introduction.}
Certifying the state of a quantum system constitutes a pivotal stage in numerous quantum information endeavors. Quantum simulation and computing represent sophisticated objectives within various quantum information pursuits necessitating validation. An established approach to tackle this issue involves making use of tomographic protocols. Conventional state tomography methods \cite{fano1957description,nielsen2002quantum,vogel1989determination} involve conducting measurements on a system in order to validate and compare the outcomes against predictions made by the Born rule. This approach is referred to as device-dependent since it assumes that measurements are perfectly characterized, a premise that is often impractical in numerous scenarios. Moreover, measurements can also be certified device-dependently, through the preparation of, in turn, perfectly characterized quantum states, introducing a form of circularity in the procedure.
	
The strongest form of device certification should then minimize the assumptions being made. The device-independent approach \cite{acin2007device,pironio2009device,colbeck2009quantum}, where a device is treated simply as a black box \cite{scarani2013device,pironio2016focus,acin2017black}, characterizes an experiment based on observed input-output measurement statistics, except for acknowledging the validity of the quantum theory. Thus, in a device-independent scenario, one can guarantee the functionality of the devices without making any assumptions about their inner workings. The feasibility of self-testing arises from the presence of non-local correlations within quantum theory. While correlations generated by classical entities remain confined to local domains, the capability to generate non-local correlations stems from the measurement of entangled states \cite{Bel64}. It's widely recognized that specific entangled states can be self-tested. This means that a classical verifier can confirm the presence of such a state among the participants by detecting the highest violation of a Bell inequality, the optimal success probability in a non-local game played by those involved, or by observing correlations that uniquely arise from measurements on that particular state. The most renowned illustration of a self-tested state is the maximally entangled qubit pair, often referred to as the singlet state. A prime method for self-testing this state involves measuring the Clauser-Horne-Shimony-Holt inequality (CHSH) to identify its maximal violation \cite{popescu1992states,summers1987bell}.
	
The concept of self-testing was first introduced by Mayers and Yao \cite{MY2004}, and since then, there has been an increasing interest in developing self-testing techniques \cite{mckague2012robust,bamps2015sum,wang2016all,kaniewski2017self,vsupic2016self,bowles2018self,kaniewski2019maximal,sarkar2021self,rai2021device,valcarce2022self,rai2022self,mckague2012robust,yang2013robust,yang2014robust,kaniewski2016analytic,baccari2020scalable,bancal2015physical}. In 2017, Coladangelo \textit{et al}. \cite{coladangelo2017all} provided a generalized scheme of self-testing for all bipartite pure entangled states with arbitrary local dimensions. It is noteworthy to mention that the majority \cite{vsupic2020self} of the presently known self-testing methods are tailored to bipartite states, and the multipartite scenario has been relatively unexplored. This is not surprising given the complexity of entanglement in the multipartite scenario. The known examples cover only the tripartite $W$ state, a class of partially entangled tripartite states \cite{wu2014robust,pal2014device}, graph states \cite{mckague2014self,baccari2020scalable} and Dicke states \cite{kaniewski2016analytic,vsupic2018self}. Recently, \v{S}upi\'{c} \textit{et al}. \cite{vsupic2023quantum} proposed a network-assistant scheme to certify all the pure entangled states. 
In the context of a network scenario, the assumption is made that the configuration entails multiple independent sources. Their scheme requires the preparation of a number of singlets that scale linearly with the number of systems, and the implementation of standard projective Bell measurements. The practical implementation of this scheme can be hindered by various realistic limitations, making them difficult to execute in experiments. 
	
	Therefore, discovering network-assist free device-independent certification schemes that are experimentally convenient, and require minimal resources and effort for practical implementation is a crucial issue of interest. In this regard, we evaluate Hardy's test of non-locality in a $n$-party scenario, which measures the highest possible violation of locality constraints permitted by quantum formalism, regardless of the system's dimension. We discover that the obtained value matches the maximum achievable violation with $n$-qubit systems. Additionally, we demonstrate that only a particular class of states can yield such a maximal value, emphasizing Hardy's test as a device-independent self-test protocol specifically designed for these states. Note that these states are genuine multipartite entangled states \cite{RWZ14}. We also, demonstrate that how our scheme is venerable against the realistic noise scenario and as an example we provide detail analysis of a tripartite system in non-ideal noise case. 
	
	The subsequent sections of this paper are structured as follows. In section \ref{sec1}, we initially present Hardy's non-locality argument, along with its extension for multipartite systems. In section \ref{sec2}, we characterize all pure $n$-qubit states capable of passing the Hardy-type non-locality test. Additionally, we derive the device-independent bound for Hardy's non-locality argument and demonstrate the self-testing property of quantum states that achieve this bound. Moving on, in section \ref{sec3}, we introduce a modified version of Hardy's test, which is experimentally feasible in a realistic noise scenario. We studied this in the context of a tripartite system. We derive the corresponding results and outline their implementation. Lastly, we summarize our work in the concluding section \ref{sec4}.

	\section{Hardy's non-locality argument}\label{sec1}
	We will first briefly discuss the Hardy paradox \cite{Har92}. Consider two parties, say Alice and Bob share a bipartite system. Both of them have a choice of two dichotomic observables $U_x$ and $D_x$, $x\in\{1,2\}$, with binary outcomes $\{+1,-1\}$. Let $P(y_1,y_2|Y_1,Y_2)$ be the joint probability that Alice and Bob have outcomes $(y_1,y_2)$ conditioning on the measurement setting $(Y_1, Y_2)$. Hardy showed that if the four conditions,
	
	\begin{equation}\label{hardy0}
		\begin{split}
			P(+1,+1|U_1,U_2)&=p>0,\\
			P(+1,+1|D_1,U_2)&=0,\\
			P(+1,+1|U_1,D_2)&=0,\\
			P(-1,-1|D_1,D_2)&=0.
		\end{split}
	\end{equation}

	are satisfied, then the resulting behavior is necessarily non-local. The degree of non-locality is characterized by the probability denoted as $P(+1,+1|U_1,U_2)$. In the context of quantum mechanics, the maximum achievable value of $P(+1,+1|U_1,U_2)$ is $\frac{-11 + 5\sqrt{5}}{2}$. The maximum can be achieved with projective measurements on a pure two-qubit state \cite{RZS12}.
	
	Now we will extend the argument for multipartite system. Consider $n$ distant parties share a multipartite system and each of them has a choice of two dichotomic observables $U_i$ and $D_i$, with binary outcomes $\{+1,-1\}$. Here, out of many possible generalizations, we consider the following one \cite{RWZ14},
	
	\begin{equation}\label{MHardyn}
		\begin{split}
			P(+1,+1,\dots,+1|U_1,U_2,\dots,U_n)&= p>0,\\
			\mbox{for~}i=1,2,\dots,n,\quad P(+1,+1|D_i,U_{i+1})&=0,\\
			P(-1,-1,\dots,-1|D_1,D_2,\dots,D_n)&=0,
		\end{split}
	\end{equation}

	with the convention $n+1\equiv 1$. Similarly to the earlier set of conditions (\ref{hardy0}), this set of conditions cannot be satisfied by any local realistic theory but can be satisfied in quantum theory. In fact, {\em only a unique pure $n$-qubit genuine entangled state satisfies all these conditions (\ref{MHardyn})} \cite{RWZ14}.

	\section{$n$-qubit states showing Hardy's non-locality}\label{sec2}
	Let us denote the eigenstates of $U_j (D_j)$ with eigenvalue $+1$ and $-1$ by $|0_j\rangle (|+_j\rangle)$ and $|1_j\rangle (|-_j\rangle)$, respectively and define $2^n+1$ product states of the $n$-qubit system $\mathcal{H}(=\mathcal{C}^{2}\otimes \mathcal{C}^{2}\otimes \dots \otimes \mathcal{C}^{2})$ of dimension $2^n$ as follows: 
	
	\begin{equation}
		\begin{aligned}
			\ket{\phi_k}&=\ket{x_1x_2\dots x_n},~ x_i\in\{0,+\},~\text{for } i=1,2,\dots n,\\ 
			& \text{and}\\
			\ket{\phi_-}&=\ket{--\dots-}, 
		\end{aligned} 
	\end{equation}
	where $\displaystyle k=\sum_{i=1}^nb_i2^{i-1}$ with $\displaystyle 	b_i=\begin{cases}
		0, \text{ if } x_i=+\\
		1, \text{ if } x_i=0
	\end{cases}$. Now, the observables $D_j$ can be expressed in terms of the eigenstates of the observables $U_j$ for each $j=1,2,\dots n$ as follows:
	
	\begin{equation}
		\begin{aligned}
			\ket{+}_j&=\alpha_j\ket{0}_j+\beta_j\ket{1}_j\\
			\ket{-}_j&=\beta_j^*\ket{0}_j-\alpha_j^*\ket{1}_j,
		\end{aligned}
	\end{equation}
	
	where
	$|\alpha_j|^2 + |\beta_j|^2 = 1$ and $0 < |\alpha_j|, |\beta_j|<1$.
	The last condition is due to the non-commutativity of $U_j$ and $D_j$. One can easily check that $\ket{\phi_-}\perp \ket{\phi_k}$ for $k=0,1,\dots, 2^{n}-2$ and $\ket{\phi_1},\ket{\phi_2},\dots,\ket{\phi_{2^n-1}}$ and $\ket{\phi_-}$ are $2^n$ linearly independent vectors of the Hilbert space $H=\mathcal{C}^{2^n}$ of dimension $2^n$. Hence, $\displaystyle \left\{\ket{\phi_-},\ket{\phi_1},\ket{\phi_2},\dots,\ket{\phi_{2^n-1}}\right \}$ forms a basis of $\mathcal{H}$ i.e., $\displaystyle \mathcal{H}=\text{Span}\left(\ket{\phi_-},\ket{\phi_1},\ket{\phi_2},\dots,\ket{\phi_{2^n-1}}\right)$.

	State $\rho$ that corresponds to conditions (\ref{MHardyn}), has to be confined to a subspace of $\mathcal{H}$, which is orthogonal to the subspace $\mathbb{S}=\text{Span} \left(\ket{\phi_-},\ket{\phi_1},\ket{\phi_2},\dots,\ket{\phi_{2^n-2}}\right)$ in view of the last two conditions of (\ref{MHardyn}). However, it is non-orthogonal to the product state $\ket{\phi_{2^n-1}}$ according to the first condition of (\ref{MHardyn}). The subspace $\mathbb{S}$ has dimension $2^n-1$, so $\rho$ must be a pure genuine n-qubit entangled state, which we denote as $\ket{\psi_n}$. Using Gram-Schmidt orthonormalization procedure one can construct an orthonormal basis $\{\ket{\phi'_i}\}_{i=0}^{2^n-1}$ for $\mathcal{H}$ from the basis $\displaystyle \left\{\ket{\phi_0},\ket{\phi_1},\dots,\ket{\phi_{2^n-2}},\ket{\phi_-}\right \}$, in which state $\ket{\psi_n}$ is its last member, with $i=2^n-1$:
	
	\begin{equation*}
		\begin{aligned}
			\ket{\phi'_0} &= \ket{\phi_-},\\
			\ket{\phi'_1}&=\ket{\phi_1},\\
			\ket{\phi'_i} &= \dfrac{\ket{\phi_i}-\sum^{i-1}_{j=0}\langle \phi'_j|\phi_i\rangle\ket{\phi'_j}}
			{\sqrt{1-\sum^{i-1}_{j=0}|\langle \phi'_j|\phi_i\rangle|^2}}, \text{ for } i=2,3,\dots, 2^n-1.
		\end{aligned} 
	\end{equation*}
	The probability $p$ in the conditions (\ref{MHardyn}), for the Hardy state, reads
	\begin{equation*}\label{value_q}
		\begin{aligned}
			p &= |\langle \psi_n |\phi_{2^n-1}\rangle|^2 \\
			&= 1-\sum_{i=0}^{2^n-2}|\langle \phi'_i|\phi_{2^n-1}\rangle|^2\\
			&=
			\dfrac{\prod_{i=1}^n |\alpha_i|^2|\beta_i|^2}{1-\prod_{i=1}^n |\alpha_i|^2}.
		\end{aligned}
	\end{equation*}
	
	The maximum probability of success of Hardy's argument (\ref{MHardyn}) for the $n$-qubits system is bounded by $p_{max}=\dfrac{t^n(1-t)^n}{1-t^n}$, where $t$ is the positive root of the polynomial $x^{n+1}-2x+1$ other than $1$. And in this optimal case {\em i.e.}, for $p=p_{max}$, $t=|\alpha_i|^2$ for $i=1,2,\dots, n$ i.e., $U_i=U$, $D_i=D$ and $\alpha_i=\alpha$, $\beta_i=\beta$ for $i=1,2,\dots, n$. Let $\ket{+}=\alpha \ket{0}+\beta \ket{1}$ and $\ket{-}=\beta^* \ket{0}- \alpha^* \ket{1}$, where $|\alpha|^2+|\beta|^2=1$, $0<|\alpha|<1$. 
	Let us denote the unique $n$-qubit Hardy state associated with this optimal case by $\ket{\psi^H_n}$. 
	
	For the new observables settings $(U,D)$ on each side, the unique Hardy state, for three-qubit system can be expressed as,
	
	\begin{equation}\label{hs3}
		\begin{split}
		\ket{\psi_3}= & c_0\ket{000}+c_1[\ket{001}+\ket{010}+\ket{100}]\\
		&+c_2[\ket{011}+\ket{101}+
		\ket{110}]+c_3\ket{111},
		\end{split}
	\end{equation}
	where $c_0=\frac{|\alpha|^3|\beta|^3}{\sqrt{1-|\alpha|^6}}$, $c_1=\frac{-\beta |\alpha|^4|\beta|}{\sqrt{1-|\alpha|^6}}$, $c_2=\frac{\beta^2|\alpha|^5}{|\beta|\sqrt{1-|\alpha|^6}}$, and $c_3=\frac{\beta^3\sqrt{1-|\alpha|^6}}{|\beta|^3}$. The success probability $p$ attains its maximal value if, and only if $|\alpha|^2 =1-|\beta|^2=
	\frac{(17+3\sqrt{33})^{2/3}-(17+3\sqrt{33})^{1/3}-2}{3(17+3\sqrt{33})^{1/3}}$. 
	
	We now state the following theorem, which is an immediate extension to a $n$-partite system of the theorem proved by Rabelo {\em et al.} \cite{RZS12} for a bipartite case.
	
	\begin{theorem}\label{deviHardy}
		In an $n$-partite Hardy test (\ref{MHardyn}), if the success probability attains its maximum value $p_{max}$, then the state of the system is equivalent up to local unitaries to $\proj{\psi^H_n}\otimes \varrho'$, where $\varrho'$ is an arbitrary $n$-partite {\em junk} state.
	\end{theorem}
	
	Here we only present the outline of the proof as the details are quite similar to the proof given in Ref. \cite{RZS12}. Let us denote the projectors of eigenstates of the operator $X$ for eigenvalues $+1$ and $-1$ as $\Pi_{+|X}=\proj{x}$ and $\Pi_{-|X}=\proj{x^{\perp}}$, respectively.
	Take two dichotomic Hermitian operators $A_1$ and $A_2$ acting on a Hilbert space $\mathcal{H}$. There exists, a decomposition of $\mathcal{H}$ as a direct sum of subspaces $\mathcal{H}^\mu$ of dimension $d \leq 2$ each, such that $A_t=\oplus_\mu A_t^\mu$ $(t=1,2)$ act within each $\mathcal{H}^\mu$ \cite{Mas06}. Therefore, $A_t^\mu=\Pi_{+|A_t^\mu}-\Pi_{-|A_t^\mu}$ and the projector $\Pi^\mu_{A_t}=\Pi_{+|A_t^\mu}+\Pi_{-|A_t^\mu}$, where each $\Pi_{\pm|A_t^\mu}$ and $\Pi^\mu_{A_t}$ act on $\mathcal{H}^\mu$, for $t=1,2$. For this notation on each side, all joint probabilities can be re-expressed as
	
	\begin{eqnarray}\label{prob_n}
		&P(a,b,\dots,c|X,Y,\dots,Z)=Tr[\rho (\Pi_{a|X}\Pi_{b|Y}\dots\Pi_{c|Z})]&\nonumber\\
		&=\sum_{\mu,\nu,\dots,\gamma}\alpha_{\mu\nu\dots\gamma}P(a,b,\dots,c|X^\mu,Y^\nu,\dots,Z^\gamma),&
	\end{eqnarray}
	where $\alpha_{\mu\nu\dots\gamma}=Tr(\rho \Pi^\mu_X \Pi_Y^\nu\dots \Pi_Z^\gamma)\geq 0$ with $\sum_{\mu,\nu,\dots,\gamma}\alpha_{\mu\nu\dots\gamma}=1$ and $P(a,b,\dots,c|X^\mu,Y^\nu,\dots, Z^\gamma)=Tr[\rho_{\mu\nu\dots\gamma}(\Pi_{a|X^\mu}\Pi_{b|Y^\nu}\dots \Pi_{c|Z^\gamma})]$ with $\rho_{\mu\nu\dots\gamma}=\frac{(\Pi^\mu_{X}\Pi^\nu_{Y}\dots \Pi^\gamma_{Z})\rho( \Pi^\mu_{X} \Pi^\nu_{Y}\dots \Pi^\gamma_{Z})}{\alpha_{\mu\nu\dots\gamma}}$. All the joint probabilities $P(a,b,\dots,c|X,Y,\dots,Z)$ will satisfy Hardy's conditions (\ref{MHardyn}) only if each decomposed probability set $P(a,b,\dots,c|X^\mu,Y^\nu,\dots,Z^\gamma)$ satisfies all the conditions $(\ref{MHardyn})$ for their associated subspace. For the maximal probability of success $P(+,+,\dots, +|U_1,U_2,\dots,U_n)$ of Hardy's argument (\ref{MHardyn}), each decomposed probability set $P(a,b,\dots,c|X^\mu,Y^\nu,\dots,Z^\gamma)$ must also attain the maximum value for its own $n$-qubit Hardy's argument,
	\begin{equation}
		\label{eq_n}\begin{aligned}
			P(+1,+1,\dots,+1|U^\mu_1,U^\nu_2,\dots, U^\gamma_n) &= p_{\mu\nu\dots\gamma}>0,&\\
			P(+1,+1|D^\mu_r,U^\nu_{r+1}) &= 0,\\
			P(-1,-1,\dots, -1|D^\mu_1,D^\nu_2,\dots, D^\gamma_n) &= 0,
		\end{aligned} 
	\end{equation}

	where $r=1,2,\dots,n$ and $U^\mu_1,D^\mu_1$, {\em etc.}, are $2\times 2$ dimensional observables. The state $\rho_{\mu\nu\dots\gamma}$, which satisfies conditions (\ref{eq_n}) must be a unique genuinely entangled pure $n$-qubit state $\proj{\chi_{\mu\nu\dots\gamma}}$ \cite{RWZ14}. Therefore, a state $\rho$, which satisfies all the conditions of (\ref{MHardyn}) is of the form,
	
	\begin{equation}\label{rho}
		\rho=\bigoplus_{i,j,k}a_{\mu\nu\dots\gamma}\proj{\chi_{\mu\nu\dots\gamma}},
	\end{equation}
	
	where coefficients $a_{\mu\nu\dots\gamma}$ are arbitrary probabilities. A state $\rho$ can lead to a maximum value of $p=p_{max}(=p_{Hardy}, say)$ if and only if all $\ket{\chi_{\mu\nu\dots\gamma}}$s' also lead to the maximal value of their corresponding $p_{\mu\nu\dots\gamma}=p_{max}$. When each $p_{\mu\nu\dots\gamma}$ reaches its own maximum $p_{max}$ then the associated {\em Hardy state} $\ket{\chi_{\mu\nu\dots\gamma}}$ must be equal to $\ket{\psi^H_n}$ for all $\mu,\nu$ and $\gamma$. Therefore, for maximum value of $p$, the state $\rho$ always satisfies
	
	\begin{eqnarray}
		&(\mathcal{O}_1\otimes\dots\otimes\mathcal{O}_n)(\rho_{12\dots n}\otimes\varrho_{1'2'\dots n'}) (\mathcal{O}^{\dag}_1\otimes\dots\otimes\mathcal{O}^{\dag}_n)&\nonumber\\
		&=\proj{\psi^H_{12\dots n}}\otimes \varrho'_{1'2'\dots n'},&
	\end{eqnarray}
	where $\mathcal{O}_i$'s are the local unitaries for $i=1,2,\dots,n$.

	\begin{theorem}\label{genuine}
		Only a genuine multipartite entangled state satisfies the Hardy-type conditions (\ref{MHardyn}) \cite{RWZ14}.
	\end{theorem}

	\section{True Hardy argument in realistic noise-tolerance scenario}\label{sec3} 
	In an ideal Hardy's argument, for the tripartite case, the constraints demand that four of the joint probabilities should be zero. However, in any real experiment, this may be very difficult to ensure. So we take a more realistic approach by considering that the constraints have the following form,
	
	\begin{equation}\label{MHardy3N}
		\begin{split}
 P(+1,+1,+1|U_1,U_2,U_3) &\geq p-\sum_{i=1}^4 \epsilon_i,\\
		\mbox{for~}i=1,2,~ P(+1,+1|D_i,U_{i+1})&\leq\epsilon_i,\\
		P(+1,+1|U_1,D_3)&\leq\epsilon_3,\\
		P(-1,-1,-1|D_1,D_2,D_3)&\leq\epsilon_4,
		\end{split}
	\end{equation}
	where $\epsilon_i \geq 0$ is some small error bound. For the sake of simplicity, we are considering $\epsilon_i = \epsilon$ for all $i$. Now the modified Hardy's argument becomes, 
	
	\begin{equation}\label{MHardy3N1}
		\begin{split}
		\textrm{Maximize:} \quad P(+1,+1,+1|U_1,U_2,U_3)&\\
		\textrm{Subject to:}\quad \mbox{for~}i=1,2,~P(+1,+1|D_i,U_{i+1})&\leq\epsilon,\\
		P(+1,+1|U_1,D_3)&\leq\epsilon,\\
		P(-1,-1,-1|D_1,D_2,D_3)&\leq\epsilon.\\
		P(abc|xyz) = Tr(\rho \: \Pi^a_x \:\Pi^b_y\: \Pi^c_z)&
		\end{split}
	\end{equation}

 One can easily verify that under local realistic model 
	 $P(+1,+1,+1|U_1,U_2,U_3) \leq P(-1,-1,-1|D_1,D_2,D_3)+P(+1,+1|D_1,U_2)+P(+1,+1|D_2,U_3)+P(+1,+1|U_1,D_3)$. So if we consider that each probability on the right side is less than equal to $\epsilon$, then $P(+1,+1,+1|U_1,U_2,U_3)\leq 4\epsilon$. Hence the local bound on Hardy’s probability of success becomes,
	$P(+1,+1,+1|U_1,U_2,U_3)\leq 4\epsilon$ with the constraints,
	\begin{eqnarray}
		\mbox{for~}i=1,2,~ P(+1,+1|D_i,U_{i+1})&\leq\epsilon,&\nonumber\\
		P(+1,+1|U_1,D_3)&\leq\epsilon,&\label{MHardy3N2}\\
		P(-1,-1,-1|D_1,D_2,D_3)&\leq\epsilon.&\nonumber
	\end{eqnarray}

\begin{figure}
		\centering
		\includegraphics[width=0.5\textwidth]{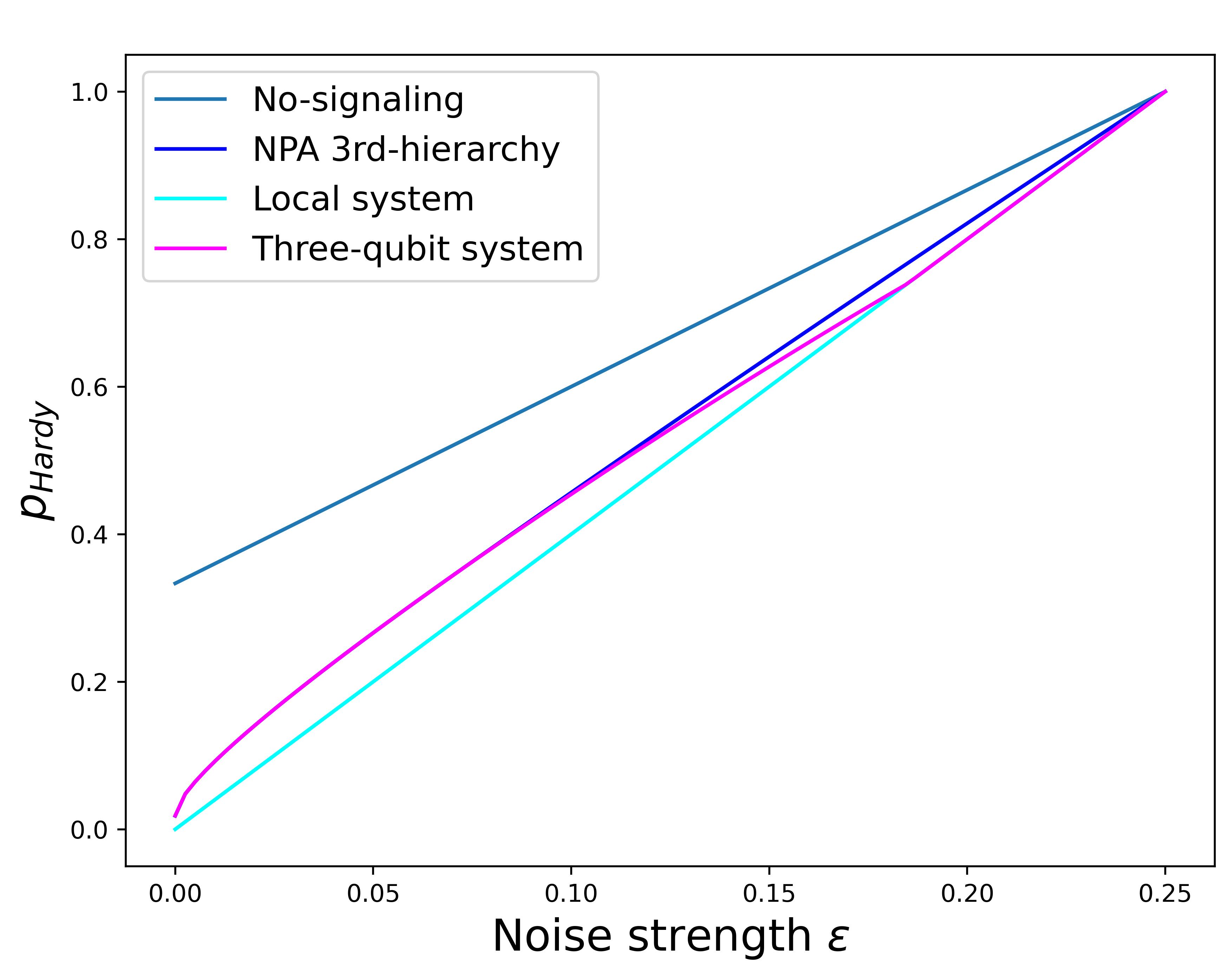}
		\caption{Plot of maximum probability of success ($p_{Hardy}$) of tripartite Hardy's argument against the noise parameter $\epsilon$ under different scenario.}
		\label{fig:3H}
	\end{figure}

	When $\epsilon \geq \frac{1}{4}$, the bound is easily satisfied, and quantum physics cannot violate it. However, for values $0 \leq \epsilon < \frac{1}{4}$, quantum physics may lead to a violation of the local bound. In the following discussion, we demonstrate that for sufficiently large values of the error bound $\epsilon$, the maximum value of the probability $\displaystyle p_{Hardy} = \max_{0 < |\alpha_1|,|\alpha_2|,|\alpha_3| < 1} P(+1,+1,+1|U_1,U_2,U_3)$ can still be achieved using pure three-qubit states and projective measurements. The analytical technique to prove the result in the ideal scenario cannot directly infer the realistic noise case. Therefore, we consider a numerical approach. In this view, we first derive the maximal quantum violation in a device-independent context, i.e., we do not make any assumptions about the Hilbert space dimension and the measurement directions. To accomplish this, we utilize semi-definite programs to obtain an upper bound on Hardy's probability, following the methodology developed by Navascu\'{e}s, Pironio, and Ac\'{i}n \cite{NPA10}. 
	Consider the set $\mathcal{Q}$ comprising quantum joint probability distributions. Instead of optimizing Hardy's probability over this quantum set $\mathcal{Q}$ directly, which can be challenging due to its complex characterization, we optimize it over a more manageable set of probabilities. This alternative set, part of an infinite hierarchy of sets denoted as $\mathcal{Q}_1 \supset \mathcal{Q}_2 \supset \dots \supset \mathcal{Q}_n \supset \dots $, is defined using semi-definite programs. Importantly, it is proven that as these sets become progressively more restrictive, they converge to the quantum set, mathematically expressed as $ \lim_{n\to\infty} \mathcal{Q}_n = \mathcal{Q}$ \cite{NPA10,DLTW08}. For various values of $\epsilon$ within the interval $0 \leq \epsilon \leq \frac{1}{4}$, we perform the optimization of Hardy's probability over the set $\mathcal{Q}_3$, while adhering to the constraints (\ref{MHardy3N2}). To derive the maximum probability of success by a quantum state (call it, quantum lower bound) we considered the following class of three-qubit states,
	\begin{equation*}
		\begin{split}
			&|\psi\rangle_{ABC} = c_{000}e^{-i(\phi+\xi+\theta)}|000\rangle +\\ &c_{001}\left(e^{-i(\phi+\theta)}|010\rangle+e^{-i(\xi+\theta)}|100\rangle+e^{-i(\phi+\xi)}|001\rangle\right)+\\
			&c_{011}\left(e^{-i\phi}|011\rangle+e^{-i\xi}|101\rangle+e^{-i\theta}|110\rangle\right)+c_{111}|111\rangle, 
		\end{split}
	\end{equation*}
	
	and the following projective measurements,
	\begin{equation*}
		\begin{split}
			U_1 &\equiv \left\{
			\begin{array}{l}
				\ket{U_1=+1} = |0\rangle,\\
				\ket{U_1=-1} = |1\rangle
			\end{array}\right\},
			\quad \\
			D_1 &\equiv \left\{
			\begin{array}{l}
				\ket{D_1=+1} = \cos(\frac{\alpha}{2}) |0\rangle + e^{i\phi}\sin(\frac{\alpha}{2}) |1\rangle,\\
				\ket{D_1=-1} = -\sin(\frac{\alpha}{2}) |0\rangle + e^{i\phi}\cos(\frac{\alpha}{2}) |1\rangle
			\end{array}\right\},
			\quad \\
			U_2 &\equiv \left\{
			\begin{array}{l}
				\ket{U_2=+1} = |0\rangle,\\
				\ket{U_2=-1} = |1\rangle
			\end{array}\right\},
			\quad \\
			D_2 &\equiv \left\{
			\begin{array}{l}
				\ket{D_2=+1} = \cos(\frac{\beta}{2}) |0\rangle + e^{i\xi}\sin(\frac{\beta}{2}) |1\rangle,\\
				\ket{D_2=-1} = -\sin(\frac{\beta}{2}) |0\rangle + e^{i\xi}\cos(\frac{\beta}{2}) |1\rangle 
			\end{array}\right\},
			\quad \\
			U_3 &\equiv \left\{
			\begin{array}{l}
				\ket{U_3=+1} = |0\rangle,\\
				\ket{U_3=-1} = |1\rangle
			\end{array}\right\},
			\quad \\
			D_3 &\equiv \left\{
			\begin{array}{l}
				\ket{D_3=+1} = \cos(\frac{\gamma}{2}) |0\rangle + e^{i\theta}\sin(\frac{\gamma}{2}) |1\rangle,\\
				\ket{D_3=-1} = -\sin(\frac{\gamma}{2}) |0\rangle + e^{i\theta}\cos(\frac{\gamma}{2}) |1\rangle
			\end{array}\right\}
		\end{split}
	\end{equation*}
	
	where $0 < \alpha, \beta, \gamma < \pi$ and $0 \leq \phi, \xi, \theta < 2\pi$.
	We proceeded to perform numerical optimization for $P(+1, +1, +1|U_1, U_2, U_3)$ across all potential states and measurement parameters while adhering to the constraints outlined in equation (\ref{MHardy3N2}). This process allowed us to establish quantum lower limits for various error thresholds ranging from $0$ to $\frac{1}{4}$. The quantum upper and lower bounds, along with the bounds imposed by locality, are depicted graphically in Figure \ref{fig:3H}. Notably, the disparities between the computed lower and upper bounds for Hardy's probability do not exceed values on the order of $10^{-6}$ for $\epsilon \lesssim 0.08$. 
	Consequently, this demonstrates that utilizing systems with higher dimensions does not confer an advantage over three-qubit systems, even when confronted with imperfections.

	The implementation of NPA-hierarchy was done in PYTHON using CVXPY and the quantum lower bound have been obtained using MATHEMATICA. The upper bound forms the blue solid line and the quantum lower bound forms the magenta solid line in Fig. \ref{fig:3H}. We have also included the no-signaling (teal solid line) and local (cyan solid line) scenarios in the Fig. \ref{fig:3H}.

	\section{Conclusion}\label{sec4}
	We have provided a self-testing scheme for genuine multipartite entangled states by employing Hardy's non-locality argument. In contrast to \v{S}upi\'{c} \textit{et al}. \cite{vsupic2023quantum} scheme our work is based on local operations and classical communications and does not depend on bipartite entanglement measurement. Also, our scheme is independent of network assistance. Hence, it is a self-testing scheme for genuine multipartite entangled states in true sense. Also, we have provided the analytic construction of unique $n$-party Hardy states for a given $n$-pair of non-commutative local observables. In addition, we have provided the device independent bound of the probability of success of generalized Hardy's argument. As an example, we have given the self-testing analysis in detail for three-party scenario both in ideal and realistic noise cases. Our result shows that the maximum probability of success of the three party Hardy argument in a device independent scenario is 0.018.
	
Hardy states hold significant potential for applications in quantum key distribution, random numbers generation and other secure quantum communication tasks due to their inherent entanglement structure. Additionally, the self-testing of these states contributes to a deeper understanding of quantum boundaries which itself enrich our fundamental knowledge of quantum theory. We believe that, this study will help in solving or suggesting some new problems in the area of quantum information theory, in general, and quantum cryptography and quantum foundation, in particular.

	\section{Acknowledgement.} Ranendu Adhikary acknowledges funding and support from ISI DCSW Project No. PU/506/PL-MISC/521. We would like to acknowledge stimulating discussions with Ashutosh Rai, Tamal Guha, Subhendu B. Ghosh, and Snehasish Roy Chowdhury.

	\bibliography{hardy}
	
\end{document}